\documentclass[a4paper]{PoS}
\usepackage{natbib}

\title{Perplexing correlations between Gamma-ray emission and parsec-scale jet orientation variations in the BL Lac object S5 0716+714}

\ShortTitle{Gamma-ray emission and jet outflow correlations in S5 0716+714}

\author{\speaker{B. Rani}$^a$, T. P.\ Krichbaum$^{a}$, A. P. Marscher$^{b}$, S. G. Jorstad$^{b}$, J. A. Hodgson$^{a}$, L. Fuhrmann$^{a}$, and J. Anton Zensus$^a$\\
{\bf On behalf of the {\it Fermi}/LAT Collaboration} \\
\llap{$^a$} Max Planck Institut f\"ur Radioastronomie, Auf dem H\"ugel 69, 53121 Bonn, Germany\\
\llap{$^b$} Institute for Astrophysical Research, Boston University, 725 Commonwealth Avenue, Boston, MA 02215, USA \\
E-mail: \email{brani@mpifr-bonn.mpg.de}}

\abstract{ 
The analysis of $\gamma$-ray flux variability along with the parsec-scale jet
kinematics suggests that the high-energy radiation in the BL Lac object S5 0716+714 has a significant correlation
with the mm-VLBI core flux density and with the local orientation of the inner jet flow.  For the first time in any blazar,
we report a significant correlation between the $\gamma$-ray flux variations and the variations in the 
local orientation of the jet outflow (position angle). We find that the $\gamma$-ray flux variations lead the 
7~mm VLBI core flux variations by 82$\pm$32~days, which suggests that the high-energy emission in S5 0716+714 
is coming from a region located 3.8$\pm$1.9~parsecs closer to the
central black hole than the ``core" seen on the mm-VLBI images. The results imply a strong physical and casual 
connection between $\gamma$-ray emission and the inner jet morphology in the source. 
}

\FullConference{12th European VLBI Network Symposium and Users Meeting\\
		 7-10 October 2014\\
		 Cagliari, Italy}

\begin{document}

\section{Introduction}
The tendency for the parsec-scale jets to change their orientation with time has been well established 
via long term VLBI studies on several AGN. 
Jet orientation variations  -- non-radial motion, helical paths of the jet features, curved jet structures, 
and variations in the direction of the inner jet flow, i.e.\ position angle or jet wobbling -- on sub-parsec 
to parsec scales are frequently  
observed in blazars [1;~2;~3]. 
The exact origin of these 
variations is not yet clear, although accretion disk precession, orbital motion of 
the accretion system, or instabilities (Magnetohydrodynamic (MHD), and/or  
Kelvin-Helmholtz (KH)) in the jet flow have all been suggested.
In some cases the inner jet position angle (PA) variations were found to 
correlate with the flux density 
variations at radio frequencies [1;~4]. 
However, a correlation between $\gamma$-ray flux and PA variations has not been reported so far. 
Here we report the observed correlation of $\gamma$-ray flux variations with 
the VLBI core flux density and the direction of the inner jet outflow in S5 0716+714.

 \begin{figure}[h]
\begin{minipage}{0.5\linewidth}
 \includegraphics[scale=0.46,angle=0, trim=130 0 120 0, clip]{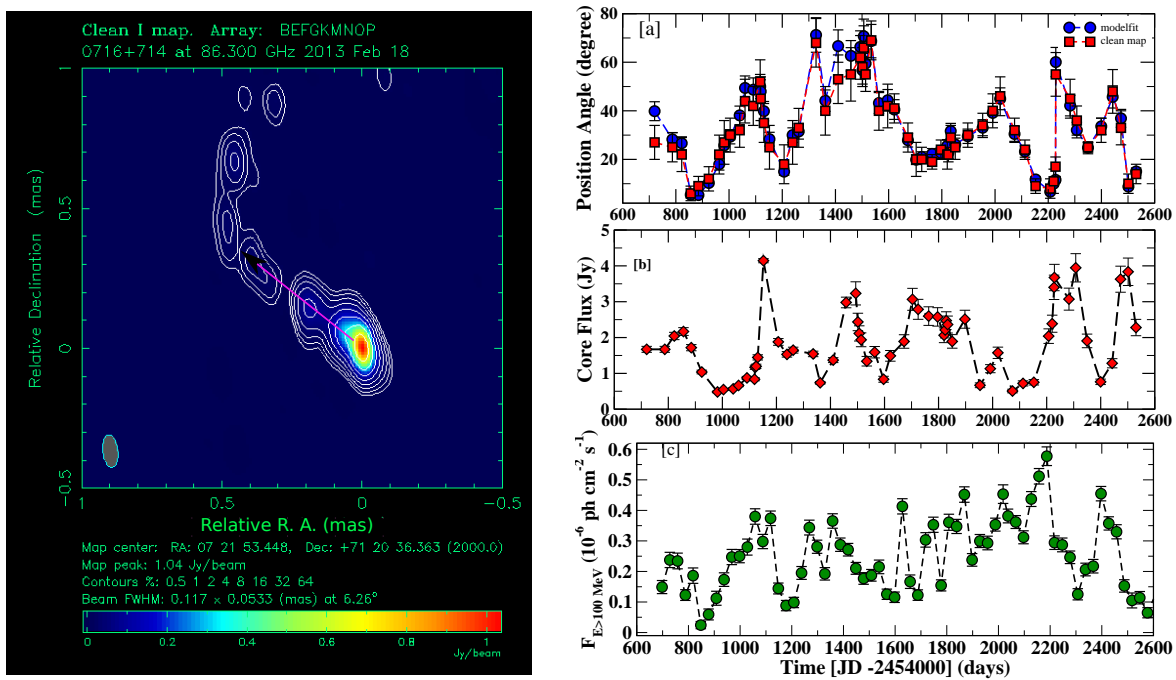} 
\end{minipage}\hfill
\begin{minipage}{0.5\linewidth}
\includegraphics[scale=0.323,angle=0,trim=0 0 0 2, clip]{PA_var.eps}
\includegraphics[scale=0.323,angle=0,trim=0 0 1.8 2, clip]{core_flx_var.eps} 
\includegraphics[scale=0.323,angle=0,trim=4 0 0 2, clip]{gamma_var.eps}
\end{minipage}\hfill
\caption{Left: Example of a 86~GHz VLBI image of S5 0716+714 observed on February 18, 2013. 
The arrow marks the orientation of inner jet outflow. Right: (a) Inner jet orientation (position angle) 
variations in the inner jet region ($\leq$0.2~mas), (b) the 7~mm VLBI core flux density variations,  and (c) 
monthly averaged $\gamma$-ray flux light curve.     }
\label{fig1}
\end{figure}

\section{Results}
For the study presented here, we used the 7~mm and 3~mm VLBI data of the source observed between August 2008 and 
September 2013. 
The inner jet orientation variations were determined taking a 
flux density-weighted average of the clean delta components at 3 times above the image noise level. In an alternative 
approach, we measured the position angle of the innermost Gaussian model-fitting. As an example a 3~mm VLBI map of the 
source is shown Fig.\ \ref{fig1} (left). 
In right panel, we plot the  position angle (PA) variations in the central region of the jet (a), and the 7~mm VLBI core flux 
density variations (b).   The $\gamma$-ray (0.1-300~GeV) data obtained by the {\it Fermi}-LAT (Large Area Telescope) 
were used to investigate the high-energy flux variations over the same time period. The details of observations and data
reduction can be found in [3]. The monthly averaged $\gamma$-ray photon flux light curve is shown in 
Fig. \ref{fig1} (c).

  \begin{figure}
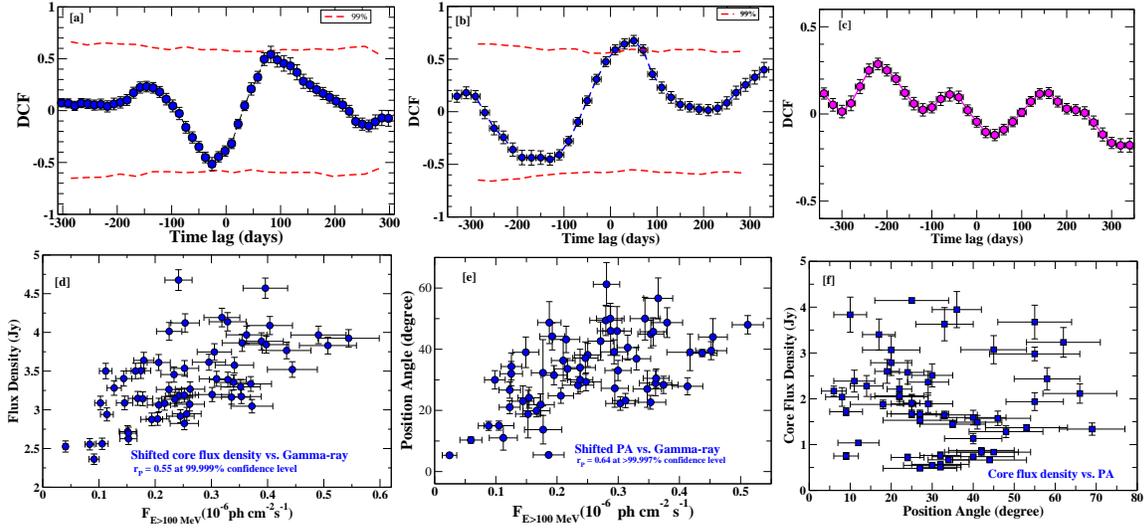

  \centering
\includegraphics[scale=0.2,angle=0,trim=0 0 0 0, clip]{dcf_gamma_core.eps}
\includegraphics[scale=0.2,angle=0,trim=0 0 0 0, clip]{dcf_gamma_pa.eps}
\includegraphics[scale=0.188,angle=0,trim=0 0 0 0, clip]{dcf_core_pa.eps}
\includegraphics[scale=0.2,angle=0,trim=0 0 0 0, clip]{gamma_core_lin.eps}
\includegraphics[scale=0.2,angle=0,trim=0 0 0 0, clip]{Gamma_PA_shift.eps}
\includegraphics[scale=0.2,angle=0,trim=-7 -10 0 0, clip]{Core_PA.eps}
  \caption{Top: Cross-correlation analysis curves -- (a) gamma-ray vs.\ core flux density, (b) gamma-ray vs.\ position angle, 
and (c) core flux density vs.\ PA. The dashed lines show the 99$\%$ confidence levels. Bottom: (d) Shifted (by 80~days) 
core flux density plotted vs.\ $\gamma$-ray photon flux, (e) shifted (by ~40 days) PA plotted vs.\ $\gamma$-ray photon flux, 
and (f) core flux density plotted vs.\ PA.   }
\label{fig2}
    \end{figure}

We used the discrete cross-correlation function (DCF)[5;~6] analysis method to investigate 
a possible correlation of $\gamma$-ray flux with core flux density and jet orientation variations; the 
results are presented in  Fig.\ \ref{fig2} (top). 
We found a significant correlation between $\gamma$-ray and core flux density variations with the 
former leading the latter by a time lag of 82$\pm$39~days (Fig.\ \ref{fig2} a). The observed time lag 
places the $\gamma$-ray emission region  upstream of the mm-VLBI core by $\geq$(3.8$\pm$1.9)~parsec 
(deprojected using a viewing angle, $\theta \leq 4.9^{\circ}$ and apparent jet 
speed $\beta_{apparent}$=10) [3]. 
In Fig.\ \ref{fig2} (d), we show the time shifted (by the observed time lag) 7~mm VLBI core flux density plotted 
vs.\ the $\gamma$-ray photon flux. A clear correlation among the two can be seen, which is confirmed by a linear
correlation analysis, yielding $r_P$ = 0.55 and 99.999$\%$ confidence level, where $r_P$ is the linear Pearson
correlation coefficient. The DCF analysis also suggests that the $\gamma$-ray flux variations correlate significantly 
with the PA variations (Fig.\ \ref{fig2} b), which is also confirmed by a linear
correlation analysis with $r_P$ = 0.64 at 99.997$\%$ confidence level. The shifted PA vs.\ $\gamma$-ray 
photon flux plot is shown in Fig.\ \ref{fig2} (e).  Since the $\gamma$-ray flux variations correlate with both 
the core flux density and PA variations, we would also expect a correlation between core flux density  and PA 
variations. The formal DCF analysis however does not reveal a significant correlation between 
core flux density and the PA variations (see Fig.\ \ref{fig2} c). We note that the absence of a significant 
correlation between the two does not rule out a weak correlation or much more complex behavior. A ring like pattern in the 
core flux density vs. PA plot (Fig.\ \ref{fig2} f) could be a hint of such a more complex behavior.

\section{Summary and Conclusions}
The observed correlations of the $\gamma$-ray photon flux variations with the inner jet orientation and 
the 7~mm core flux density variations  can be interpreted as a moving shock
propagating down a relativistic jet with non-axisymmetric pressure
or density gradients/patterns, or a shock moving in a bent
jet. Alternatively, instability patterns moving downstream and passing 
the two emission regions at different viewing angles, or even a rotation of the 
helical jet around its own z-axis, would cause very similar variations [3]. 
Correlated variations can be expected if the two emission regions share 
same boosting cone. A correlated variation between the core flux density 
and the inner jet orientation is expected in a simple geometrical interpretation. 
The absence of such a correlation suggests that there are some unknown 
factors (e.g.\ opacity) in addition to geometry which suppress the core flux density and jet 
orientation correlation. High-frequency VLBI monitoring with denser time sampling
would be required to better understand the jet morphology and it's relation to the 
$\gamma$-ray emission.

\smallskip
\smallskip

\noindent 
{\bf Acknowledgments.}
{\footnotesize
The $Fermi$-LAT Collaboration acknowledges support from a number of agencies and institutes for both 
development and the operation of the LAT as well as scientific data analysis. These include NASA and 
DOE in the United States, CEA/Irfu and IN2P3/CNRS in France, ASI and INFN in Italy, MEXT, KEK, and JAXA 
in Japan, and the K.~A.~Wallenberg Foundation, the Swedish Research Council and the National Space Board 
in Sweden. Additional support from INAF in Italy and CNES in France for science analysis during the 
operations phase is also gratefully acknowledged. This study makes use of 43 GHz VLBA data from the 
VLBA-BU Blazar Monitoring Program (VLBA-BU-BLAZAR; http://www.bu.edu/blazars/VLBAproject.html). 
This research has made use of data obtained with the Global Millimeter VLBI Array (GMVA), which consists of 
telescopes operated by the MPIfR, IRAM, Onsala, Mets\"ahovi, Yebes and the VLBA. The data were correlated at 
the correlator of the MPIfR in Bonn, Germany. The VLBA is an instrument of the National Radio Astronomy 
Observatory, a facility of the National Science Foundation operated under cooperative agreement by Associated 
Universities, Inc. }

\providecommand{\href}[2]{#2}\begingroup\raggedright\endgroup

\end{document}